\documentclass[a4paper,12pt]{article}
\usepackage{epsf,amsmath}
\usepackage{indentfirst}
\usepackage{amssymb}
\usepackage{hhline}
\usepackage{array,tabularx}
\def\@biblabel#1{#1.}
\makeatother

\begin{document}

\title{\Large\bf Magnetars, Gamma-ray Bursts, and Very Close Binaries}

\author{A. I. Bogomazov\footnote{a78b@yandex.ru}, S. B. Popov\footnote{sergepolar@gmail.com} \\
\small\it Sternberg Astronomical Institute, Moscow State
University, \\ \small\it Universitetski pr. 13, Moscow, 119992,
Russia \\
\small Astronomy Reports, volume 53, no. 4, pp. 325-333 (2009)}

\date{\begin{minipage}{15.5cm} \small
We consider the possible existence of a common channel of
evolution of binary systems, which results in a gamma-ray burst
during the formation of a black hole or the birth of a magnetar
during the formation of a neutron star. We assume that the rapid
rotation of the core of a collapsing star can be explained by
tidal synchronization in a very close binary. The calculated rate
of formation of rapidly rotating neutron stars is qualitatively
consistent with estimates of the formation rate of magnetars.
However, our analysis of the binarity of newly-born compact
objects with short rotational periods indicates that the fraction
of binaries among them substantially exceeds the observational
estimates. To bring this fraction into agreement with the
statistics for magnetars, the additional velocity acquired by a
magnetar during its formation must be primarily perpendicular to
the orbital plane before the supernova explosion, and be large.
\end{minipage} } \maketitle \rm

\section{Introduction}

The nature of magnetars and gamma-ray bursts (GRBs) is a hotly
debated topic in astrophysics. Some standard models for these
objects and their origin exist, but alternative models are still
discussed, and a number of problems remain in the standard models.

Magnetars \cite{duncan1992a,woods2006a} are neutron stars whose
activity is related to the dissipation of magnetic energy, which
distinguishes them from other neutron stars. The observational
manifestations of neutron stars may be connected with the release
of potential energy of infalling matter (accreting sources),
rotational energy (radio pulsars), or thermal energy (cooling
neutron stars, such as compact objects in supernovae remnants and
the so-called ``Magnificent Seven'' --- nearby, single,
radio-quiet cooling neutron stars). Presumed magnetars (soft gamma
repeaters, SGRs, and anomalous X-ray pulsars, AXPs) display energy
releases that exceed the rate of dissipation of rotational energy
or stored heat, and accretion is not a possible source of their
activity. Magnetars display large magnetic fields of $\sim
10^{14}$~G. Their periods of rotation are roughly 2-12 s. There
exist ordinary radio pulsars with comparable fields and periods;
i.e., magnetars are not specified only by their strong magnetic
fields. It is important that the energy of the field is being
dissipated, and this source dominates in the object's luminosity.
The origin of the fields in strongly magnetized neutron stars of
all types remains unclear. In the standard model suggested by
Duncan and Thompson \cite{duncan1992a} and its modifications, this
field is generated in a proto-neutron star by the dynamo
mechanism, which requires very rapid initial rotation of the
neutron star (with periods shorter than several ms).

The standard model for long cosmic GRBs (see, for example,
\cite{woosley1993a}-\cite{bogomazov2008a} and the reviews
\cite{meszaros2006a}) invokes the collapse of a massive star, with
the formation of a black hole and a massive surrounding disk. The
formation of a disk with the necessary parameters requires rapid
rotation of the collapsing core.

As we can see, in standard models for the formation of both
magnetars and GRBs, the core of the progenitor star rotates very
rapidly. Some studies (see, for example, \cite{mqt2008, kb2007}
and references therein, as well as the text below) have considered
models in which GRBs (or at least some of them) are directly
connected with the formation of a rapidly rotating magnetar. The
requirement of rapid initial rotation is a real challenge for
descriptions of the evolution preceding the collapse, since the
cores of massive stars probably appreciably decelerate their
rotation in the course of their evolution (see, for example,
\cite{tutukov2003b}-\cite{meynet2005a}). This contradiction can be
resolved if the presupernova was in a close binary system, where
its evolution can result in an overspeeding of star rotation due
to accretion or tidal synchronization of the components.More
exotic mechanisms for untwisting forming neutron stars have also
been discussed \cite{john2006a}, but they are related to details
of themechanism for the supernova explosion, which remain not
fully clarified.

Here, we analyze the hypothesis that there exists a single channel
for the formation of GRBs and magnetars. The evolutionary track of
a binary system results in the appearance of a pre-supernova with
a rapidly rotating core. If the remnant of the evolution is a
neutron star, a magnetar is formed. If the pre-supernova was
sufficiently massive, its remnant will be a black hole, and a GRB
is observed. An indirect argument in support of this scenario (and
a motivation for our study) is the fact that, to order of
magnitude, the fraction of magnetars among neutron stars (about
10\% \cite{gill2007a, ferrario2008a}) coincides with the fraction
of GRBs among supernova outbursts (about 3\%; see, for example,
\cite{soderberg2007a}), that are considered to end in the
formation of a black hole. In the studies
\cite{metzger2007a,bucciantini2007a} (see also references therein
to earlier studies carried out by this group), a mechanism for a
GRB that is directly connected with the formation of a magnetar is
described in detail. The evolutionary channel discussed here can
certainly also result in such events.

To analyze the formation of magnetars, we must also take into
account the fact that all $\approx$15 objects of this type known
are single. Since no explicit strong selection favoring the
detection of single magnetars is known, this restriction must be
additionally taken into consideration (it was first discussed in
\cite{popov2006a}).

\section{Population synthesis}

Since the principles of the ``Scenario Machine'' have been
described many times before, here, we will only note the
parameters of the evolutionary scenario adopted as free parameters
in solution of our problem. A detailed description of the code can
be found in \cite{lipunov1996a}-\cite{scm2}. The
population-synthesis technique is also described in \cite{pp2007}.

For each set of parameters of the evolutionary scenario, we
carried out a population synthesis for $10^6$ binaries. The rates
of events and the numbers of objects in the Galaxy are given
assuming that all stars are binaries. As free parameters of the
scenario, we adopted the kick velocity acquired during the
formation of neutron stars and the mass-loss rate for
non-degenerate stars.

\subsection{The Kick Acquired in a Supernova Explosion}

In our calculations we supposed that a neutron star may acquire
some additional velocity $v$ during the supernova explosion in
which it is formed (see, for example, \cite{wang2005a} and
references therein). Here, we use two versions for the
distribution of the speed and direction associated with this kick
velocity.

In the first, the kick velocity is random and distributed
according to a Maxwellian function:

\begin{equation}
\label{Maxwell} f(v)\sim \frac{v^2}{v^3_0}e^{-\frac{v^2}{v_0^2}},
\end{equation}

\noindent where $v_0$ is a free parameter.

In the second case, the distribution of the kick velocity is a
$\delta$ function; i.e., all neutron stars acquire the same kick
velocity $v_0$.

The kick velocity was considered to be either uniformly directed,
or to be directed along the rotational axis of the neutron star.
The latter case is associated with the fact that the direction of
the rotational axis essentially provides a preferred direction.
First, in the course of a prolonged kick, averaging about this
axis will occur, and the resulting velocity will be directed along
it \cite{lcc2001}. Second, the magnetic field generated by a
dynamo mechanism will also be oriented primarily along the
rotational axis, while, in some models, the appearance of the kick
is due to asymmetrical neutrino radiation in strong magnetic
fields \cite{lcc2001}. Third, in the magneto-rotational model of a
supernova explosion, the kick velocity is primarily directed along
the rotational axis \cite{abkm05}. Finally, in a number of models
in which the kick is associated with the development of
instabilities at the supernova stage, the rotational axis also
represents a preferred direction, and calculations reveal a
correlation between the direction of the kick velocity and this
axis \cite{setal2006}. Observations of radio pulsars present some
very strong arguments suggesting that the directions of the kick
velocity and rotational axis coincide \cite{nr2007}-\cite{r2007},
while the rotational axis in binaries may be expected to be
perpendicular to the orbital plane.

Note that the kick may not only disrupt, but also bind some
systems, which would have decayed without the kick due to the
large mass loss in the supernova explosion. However, in most
cases, increasing the kick velocity acquired by the neutron star
during its formation decreases the probability that a binary will
be preserved.

\subsection{Stellar Wind}

Here, we use two evolutionary scenarios (A and C) that differ in
the mass-loss rate for non-degenerate stars. The stellar wind is
an important evolutionary parameter, since it specifies the masses
of the remnants of the stellar evolution and the semi-major axes
of binary systems.

Evolutionary scenario A features a classical weak stellar wind
(see \cite{Lipunov1996}, as well as evolutionary scenario A in
\cite{scm2}). On the main sequence and in the supergiant stage, a
star loses no more than 10\% of its mass in each of these stages,
and in the Wolf-Rayet stage it loses 30\% of its initial mass.

Evolutionary scenario C \cite{scm2} features a higher mass-loss
rate. In each stage of its evolution, a star fully loses its
envelope, which may mean the loss of more than half its initial
mass by the end of its evolution.

\subsection{Angular Momentum of a Non-degenerate Core}

The main basis for the model considered here is the acceleration
of the rotation of a pre-supernova core in a close binary due to
tidal synchronization of the rotation of its companion. In this
case, we can formally assume that the synchronization occurs just
before the collapse. The period of axial rotation of the newborn
neutron star should then be approximately $10^{-6}$ of the orbital
period of the binary at the time of the supernova explosion, since
the radius of the core, whose mass before the collapse is
approximately $1.5 M_{\odot}$, is $\sim 10^9$ cm
\cite{tutukov1988a}, while the characteristic size of the neutron
star is $\sim 10^6$ cm. We assume that the rotational angular
momentum of this core is conserved during the collapse.

However, the lifetime of the core of a star in the carbon-burning
stage is $\sim 10^4$ years \cite{tutukov1988a}, while, as a rule,
the time for the tidal synchronization of components is no less
than $\sim 10^4$ yrs, even in very close systems \cite{zahn2008a}.
Thus, we can take the synchronization of the axial rotation of the
core with the orbital rotation of the companion to occur at the
carbonburning stage, or at the end of the helium-burning stage
\cite{meynet2005a}. During the stage of the burning of carbon and
subsequent elements, the period of axial rotation of the core is
shorter than the orbital period of the system. In this case, the
rotational period of the newborn neutron star will be
approximately $10^{-8}$ of the orbital period of the binary when
the components' rotation becomes synchronized: the radius of a CO
core with an approximate mass of $1.5 M_{\odot}$ is $\sim 10^{10}$
cm \cite{tutukov1988a}, while the characteristic size of the
neutron star is $\sim 10^6$ cm. It is assumed that the rotational
angular momentumis conserved the core has possessed at the onset
of the carbon-burning stage.

We decided to call a magnetar a neutron star that has originated
in a close binary, with an initial period of axial rotation that
does not exceed 5 ms. Such rapid rotation should make it possible
to increase the magnetic field substantially due to the dynamo
mechanism. Consequently, if we suppose that the synchronization of
the rotation occurs before the collapse, then the maximum orbital
period at the epoch when the orbital rotation of the components
becomes synchronized with the rotation of the core of the future
collapsar should not exceed $\sim 10$ days.

If the formally defined rotational period of the newborn neutron
star is shorter than 0.001 s (the limiting minimum rotational
period for neutron stars), we consider the period of the neutron
star to be equal to this value. It seems to us that the excess
angular momentum may lead to additional peculiarities of supernova
explosions in such close binaries.

\subsection{Other Parameters of the Evolutionary Scenario}

In this section, we present some parameters of the evolutionary
scenario that are not yet known accurately, and so can be adopted
as free parameters in the population synthesis carried out using
the ``Scenario Machine''. The maximum mass of the neutron star
(the Oppenheimer-Volkov limit) that can be attained via accretion
is taken to be $M_\mathrm{OV}=2.0 M_{\odot}$, and the initial
masses of the young neutron stars to be distributed randomly in
the interval $1.25-1.44 M_{\odot}$.

We assume that main-sequence stars with initial masses in the
range $10$ -- $25$ complete their evolution as neutron stars.
There is some evidence that the progenitors of magnetars are the
most massive stars among those forming neutron stars
\cite{muno2006a}; however, this is not a firm fact for all
magnetars, and we do not take this possibility into account. It
would be important to distinguish supernovae in which magnetars
are formed (see, for example, \cite{maeda2007a}), but this
question is likewise unclear.

Main-sequence components of close binaries that increase their
masses as a result of mass exchange until their values appear in
the above interval were also added to the progenitors of neutron
stars. More massive stars evolve into black holes, and less
massive stars to white dwarfs. In our present calculations, we
assumed a uniform (flat) distribution of component-mass ratios for
the initial binaries \cite{tutukov1988a} and zero initial
eccentricity for their orbits. We also adopted a flat distribution
of the initial semi-major axes of the binaries, $d(\lg
a)=\mbox{const}$ in the interval $10-10^6 R_{\odot}$. The
efficiency of the mass loss at the common-envelope stage is
described by the parameter $\alpha_\mathrm{CE}=\Delta
E_\mathrm{b}/\Delta E_\mathrm{orb}=0.5$, where $\Delta
E_\mathrm{b}=E_\mathrm{grav}-E_\mathrm{thermal}$ is the binding
energy of the ejected matter of the envelope and $\Delta
E_\mathrm{orb}$ the decrease in the orbital energy of the system
as its components approach \cite{tutukov1988a,heuvel1994a}.

\section{Results}

We note first that the synchronization of the core just before the
collapse\footnote{ The period of axial rotation of a newborn
neutron star is $10^{-6}$ of the orbital period before the
collapse. We consider a magnetar to be a neutron star whose
rotational period does not exceed 5 ms. } is essentially
incompatible with the available observational data. The rate of
formation of magnetars under this assumption would be $\sim
10^{-5}$ per year, which is more than two orders of magnitude
below the lowest observational estimates (one per several hundred
years). The binarity of magnetars if the synchronization of the
core rotation occurs just before the collapse begins to differ
substantially from unity only for very high kick velocities ($\sim
10^3$ km/s), which do not seem likely. Therefore, we consider
further only the possibility that the core maintains the
rotational angular momentum it possesses at the onset of the
carbon-burning stage.

\begin{figure}
\hspace{0cm} \epsfxsize=0.9\textwidth\centering \epsfbox{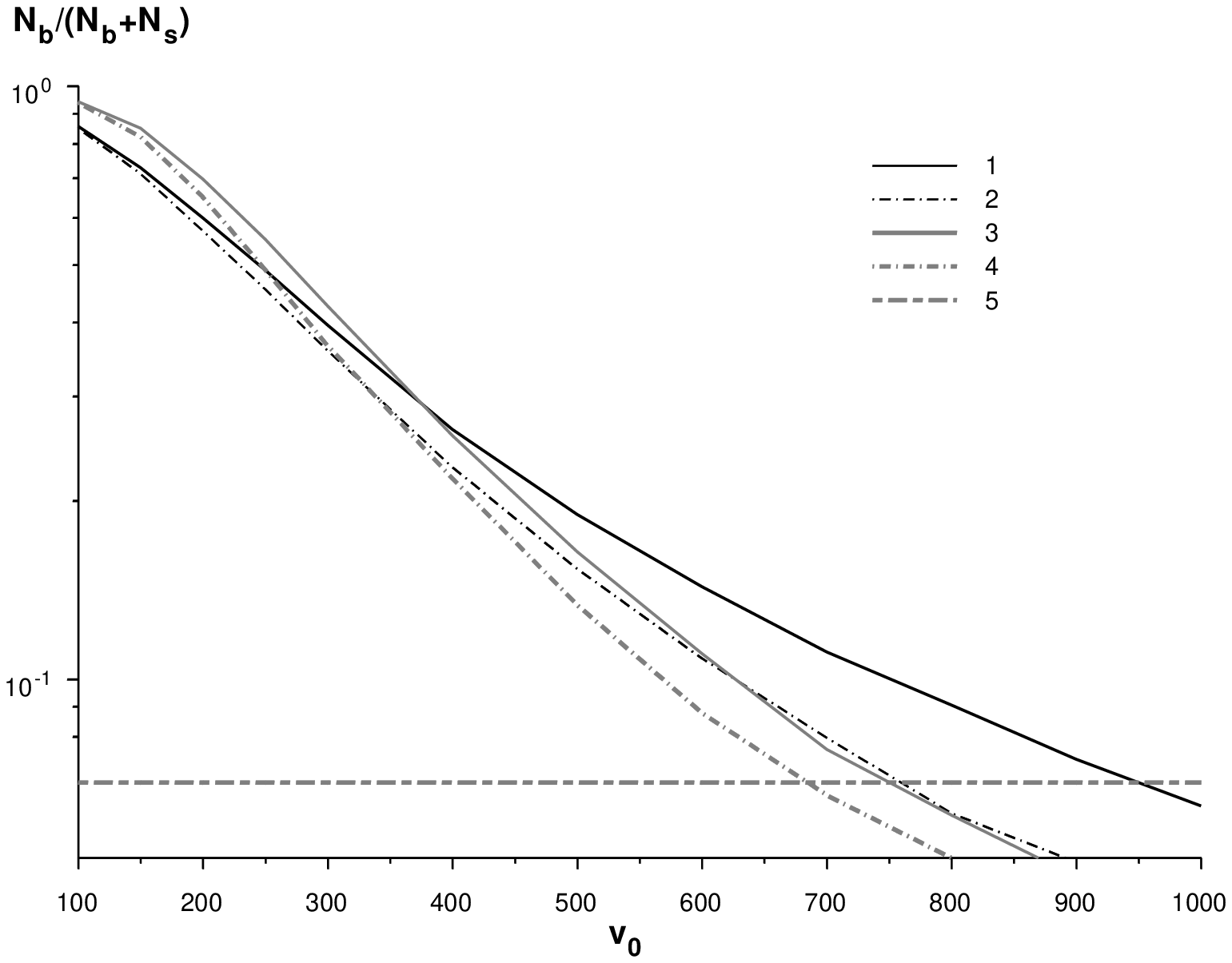}
\vspace{0cm}\caption{ The fraction of binary neutron stars formed
in very close binaries as a function of the characteristic kick
velocity $v_0$ acquired during the formation of a neutron star.
The kick velocity is distributed according to (\ref{Maxwell}). The
vertical axis plots $N_\mathrm{b}/(N_\mathrm{b}+N_\mathrm{s})$,
where $N_\mathrm{s}$ is the number of singular neutron stars and
$N_\mathrm{b}$ the number of neutron stars in binaries that
originated in the course of the computations. The numbers on the
graph denote the curves calculated for the following scenarios:
({\it 1}) equiprobable directions for the kick, type A
evolutionary scenario; ({\it 2}) equiprobable directions for the
kick, type C evolutionary scenario; ({\it 3}) kick always
perpendicular to the plane of axial rotation of the star, type A
evolutionary scenario; ({\it 4}) kick always perpendicular to the
plane of axial rotation of the star, type C type evolutionary
scenario; ({\it 5}) upper limit for binarity according to
observations ($\approx 1/15$). } \label{r1}
\end{figure}

\begin{figure}
\hspace{0cm} \epsfxsize=0.9\textwidth\centering \epsfbox{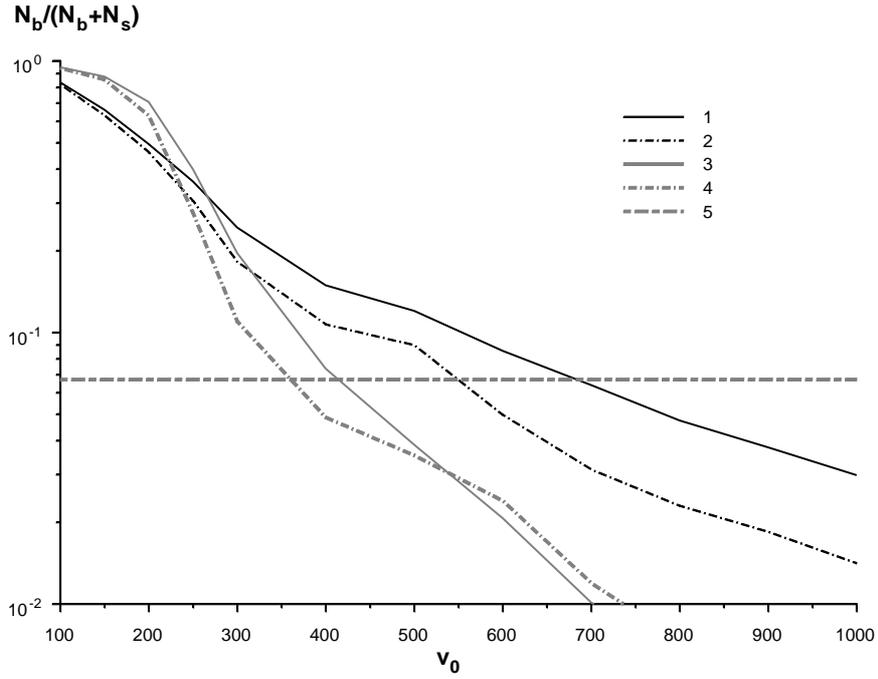}
\vspace{0cm}\caption{ Same as Fig. \ref{r1} for a
$\delta$-function distribution for the kick velocity. The numbers
on the graph denote the curves calculated for the following
scenarios: ({\it 1}) equiprobable kick-velocity direction, type A
evolutionary scenario; ({\it 2}) equiprobable kick-velocity
direction, type C evolutionary scenario; ({\it 3}) kick velocity
directed along the rotational axis of the star, type A
evolutionary scenario; ({\it 4}) kick directed along the
rotational axis of the star, type C evolutionary scenario; ({\it
5}) upper limit for binarity according to observations ($\approx
1/15$). } \label{r2}
\end{figure}

\begin{figure}
\hspace{0cm} \epsfxsize=0.9\textwidth\centering \epsfbox{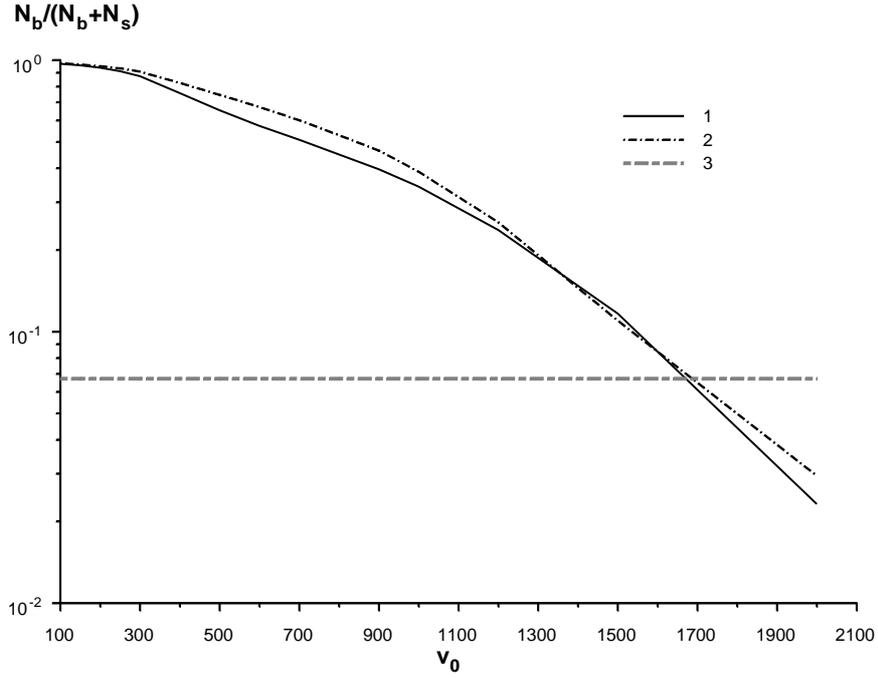}
\vspace{0cm}\caption{ Same as Fig. \ref{r1} for a
$\delta$-function kick-velocity distribution with the kick
directed along the rotational axis of the star; the absolute value
of the kick also depends on the initial rotational period of the
young neutron star (see text for the details). The numbers on the
graph denote the curves calculated for the following scenarios:
({\it 1}) type A evolutionary scenario, ({\it 2}) type C
evolutionary scenario, ({\it 3}) upper limit for binarity
according to observations ($\approx 1/15$). } \label{r3}
\end{figure}

\begin{figure}
\hspace{0cm} \epsfxsize=0.9\textwidth\centering \epsfbox{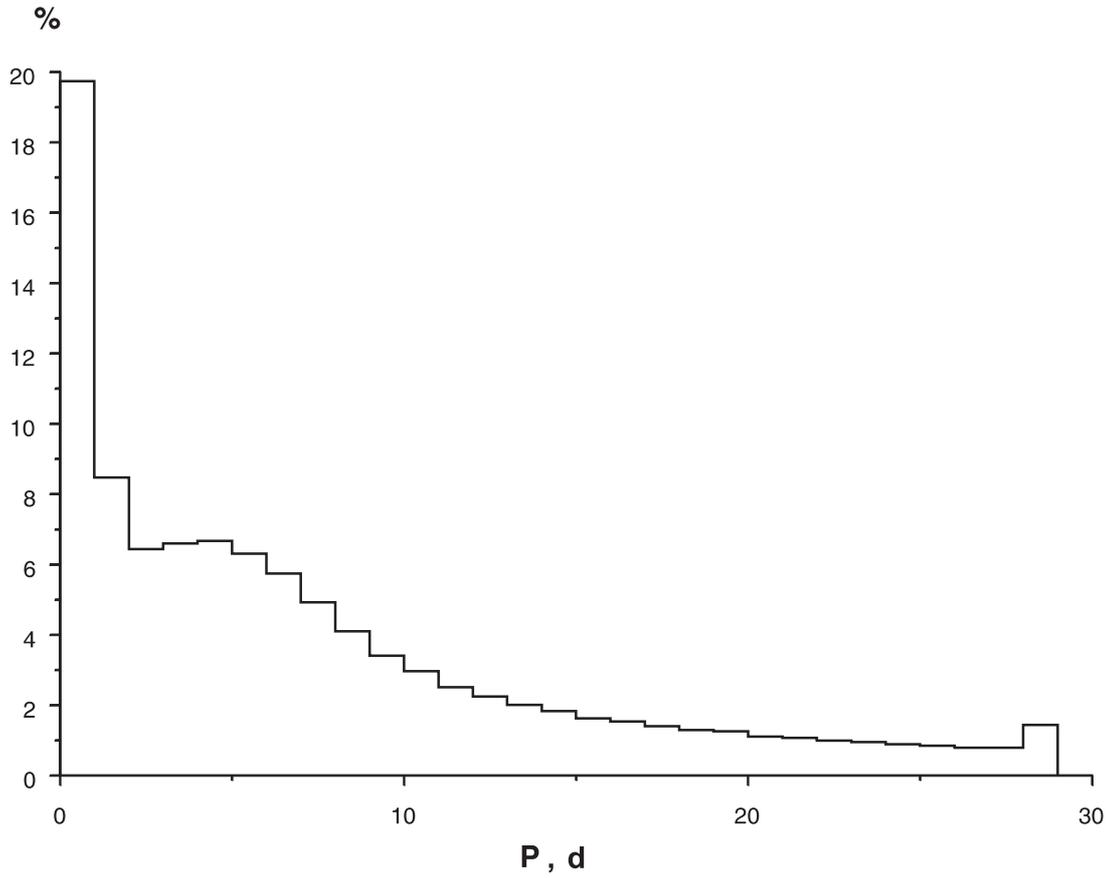}
\vspace{0cm}\caption{ Distribution of orbital periods just before
the collapse in the systems in which neutron stars originate. If a
neutron star originates in a disrupted system, the orbital period
at the time of disruption is taken into account. The type A
evolutionary scenario is adopted (a weak stellar wind). }
\label{period}
\end{figure}

According to our computations, the rate of formation of rapidly
rotating neutron stars (supposed to be magnetars) is approximately
one per 400-500 yrs. This agrees with empirical estimates for the
rate of formation of these objects (see the analysis and detailed
discussion in \cite{gill2007a}). Figures \ref{r1}-\ref{period}
present the results of our computations. All the curves in Figs.
\ref{r1}-\ref{r3} assume that the angular momentum due to the
axial rotation of the CO core is maintained, and, just before the
collapse, the pre-supernova core rotates with a period shorter
than the orbital period of the system at the time of the
supernova. The maximum orbital period of the binary at the epoch
when the rotation of the components becomes synchronized is 10
days.

We can clearly see in Fig. \ref{r1}, that the binarity of
magnetars remains too high if the magnitude of the kick velocity
displays a Maxwell distribution, regardless of the other
parameters of the evolutionary scenario, up to very high values
$v_0$ ($>700$ km/s). The curves in Fig. \ref{r2} assume that the
kick velocity has some specific value $v_0$ (it is a $\delta$
function). This makes it possible to reduce the predicted binarity
of magnetars. The presence of a strong stellar wind (evolutionary
scenario C) decreases the binarity (curves 2 and 4 in Figs.
\ref{r1} and \ref{r2}, and curve 2 in Fig. \ref{r3}) compared to
scenarios with a weak wind. If the kick is uniformly directed and
the stellar wind is weak (evolutionary scenario A), consistency
with observations can be reached for $v_0\approx 700$ km/s (curve
1 in Fig. \ref{r2}), even if the additional kick is represented by
a $\delta$ function. With curves 3 and 4 in Fig. \ref{r2} and
moderately large kick velocities (350-450 km/s), it is possible to
reach a level of binarity for magnetars that corresponds to the
current observational data (the fraction of binaries is $<1/15$).
This becomes possible if the kick is primarily directed along the
rotational axis of the forming neutron star, perpendicular to the
orbital plane of the binary at the time of the supernova (curves 3
and 4). A strong wind (curves 2 and 4) also decreases the
predicted binarity of magnetars, but having the primary direction
of the kick along the rotational axis of the newborn neutron star
is preferably to achieve consistency between the computations and
observations (curve 4).

The curves in Fig. 3 assume that the kick is directed along the
rotational axis of the forming neutron star and that the kick is
represented by a $\delta$ function, but also depends on the
orbital period at the time of the supernova as $v=v_0\cdot
0.001/P_\mathrm{NS}$, where $0.001\mbox{ s }\le P_\mathrm{NS}\le
0.005\mbox{ s}$ is the period of the forming neutron star. The
orbital period is restricted to approximately 10 days. The need
for this computation stems from the following consideration. Our
model assumes that the very strong magnetic field of the magnetar
is generated as a result of the collapse of a very rapidly
rotating core. In our present study, the maximum and minimum
periods of magnetars differ by a factor of five. The kick velocity
at the time of the collapse may depend on the magnetic field,
while the predicted binarity of magnetars may depend on the
orbital period at the time of the supernova. Figure \ref{r3},
shows that curve 3 corresponds to observational data starting from
$v_0=1700$ km/s. For systems whose orbital periods at the time of
tidal synchronization are, for example, 5 days, the kick velocity
will not exceed 400 km/s, while the largest kick will be obtained
by neutron stars forming in binaries with orbital periods
$\lesssim 1$ day. This graph may provide evidence against the
basic model considered here.

We can see from Fig. \ref{period} that the maximum orbital periods
of binaries in which magnetars can form can be decreased to
several days; however, they cannot be shorter than a day, since
the rate of formation of rapidly rotating neutron stars would then
be too low.

The potential companions in hypothetical binaries containing
magnetars are also of interest. Most of these are main-sequence
stars (49\%) and black holes (46\%). The remaining 5\% are roughly
equally divided among white dwarfs (2\%), Wolf-Rayet stars (1\%),
stars filling their Roche lobes (0.7\%), helium stars filling
their Roche lobes (the BB stage), hot white dwarfs (0.7\%), and
neutron stars (0.6\%).

\section{Discussion}

\subsection{Alternatives}

We were not able to obtain a sufficient number of single magnetars
in our evolutionary scenario for binaries without making
additional assumptions, for example, about the kick (recoil)
velocity at the time of the supernova. This could be considered
indirect argument against the generation of the magnetar magnetic
field in rapidly rotating newborn neutron stars. In this
connection, we will recall and briefly discuss alternatives to the
considered scenario.

\subsubsection{Relict magnetic field}

One currently popular hypothesis suggests that the fields of
magnetars are formed in the collapses of stellar cores with
magnetic-flux conservation, when the progenitor star has a
sufficiently strong magnetic field (see \cite{ferrario2008a} and
references therein). Some observations suggest that magnetars are
related to the most massive stars among those giving rise to
neutron stars \cite{muno2006b}. According to some observational
data, about a quarter of these massive stars display sufficiently
strong magnetic fields (see \cite{petit2008a} and references
therein). In addition, studies of supernova remnants related to
magnetars have not revealed any direct signs of intense energy
release that could have been connected with the presence in them
in the past of rapidly rotating neutron stars with magnetic fields
\cite{vink2006a}. These studies can be considered indirect
arguments against the hypothesis that the magnetar fields were
generated in the course of a collapse. Simple population estimates
\cite{ferrario2008a} indicate that current estimates for the rate
of formation of magnetars can be explained in this hypothesis.

However, there are serious objections against this hypothesis,
some of them recently summarized by Spruit \cite{spruit2008a}. The
simplest is that, even if there exists a strongly magnetized
massive star, only 2\% of its cross section (which is important
when calculating the field during a collapse with flux
conservation) will be included in the compact object.

The rate of formation of magnetars remains very uncertain. Recent
detections of transient anomalous X-ray pulsars
\cite{gelfand2007a}-\cite{gavriil2008a} and the detection of a new
source of repeating GRBs \cite{holland2008a,barthelmy2008a}
suggest that the number of magnetars may exceed previous
estimates. If true, this may raise the problem of the lack of
sufficiently strongly magnetized massive stellar progenitors able
to provide the high formation rate of magnetars. Finally, studies
of stellar magnetic fields are able to probe only the surface
fields. A compact object emerges from the stellar core, whose
field is unknown.

Thus, to explain the origin of the fields of magnetars, it is
difficult to get away without some mechanism for its generation.
All the possible mechanisms \cite{spruit2008a} use the rotational
energy of a newborn neutron star or collapsing core in some way;
i.e., the question of what makes 10\% to several tens of per cent
of cores of massive stars rotate rapidly just before their
collapse remains open.

\subsubsection{Other channels of evolution in binaries}

A more optimistic scenario than ours is considered in
\cite{popov2006a}. While here we consider only stars that rotate
rapidly just before the collapse, several possible means for
spinning up stars in binaries are suggested in \cite{popov2006a},
neglecting the possible subsequent deceleration of the rotation.
It is not surprising that this led, first, to a substantially
higher rate of formation of rapidly rotating neutron stars, and
second, to a fraction of preserved binaries that was much lower.
Unlike the channels considered in \cite{popov2006a}, in the
scenario considered above, rapidly rotating neutron stars
originate only in very compact systems, and, in addition, the
exploding star is often less massive than its companion.

There are a number of objections against the optimistic suggestion
that a normal star spun up by accretion in an early stage of its
evolution, or an object formed as a result of a merger, can
maintain the high angular momentum of its core until its collapse.
For example, three processes that can transfer the rotational
angular momentum from one layer to another are considered in
\cite{meynet2005a} convection, diffusion (shear diffusion), and
meridional circulation. Convection tends to make the angular
velocity constant, thereby transferring angular momentum from
inner to outer layers of the convective zone. Diffusion weakens
differential rotation and also transfers angular momentum
outwards. Meridianal circulation can transfer angularmomentumboth
outward and inward in the star. Mass loss affects the angular
momentum of the core in part indirectly, since it influences the
rotational angular velocity of the star and the angularvelocity
gradient inside the star. The most important conclusions related
to the evolution of the rotation listed in \cite{meynet2005a} are
as follows (see also references therein):

\begin{itemize}

\item The angular momentum of the star decreases during the star's
evolution up to the supernova explosion.

\item The largest loss of angular momentum occurs on the main
sequence.

\item After helium burning has finished, convection transfers
angular momentum from inner parts of the convective zone to its
outer layers, without involving the core; therefore, the
rotational angular momentum of the core at the end of the
helium-burning stage can be taken as a fairly reliable estimate of
the rotational angular momentum of the collapsing core.

\end{itemize}

\subsection{Possible Observational Manifestations
of Magnetars in Binary Systems}

In our scenario, with an isotropic kick, an appreciable fraction
of magnetars remain in bound binaries. In this connection, we
should discuss possible manifestations of close binaries with
magnetars.

Since the magnetic fields of magnetars apparently rapidly decrease
to values typical for common radio pulsars (see, for example,
\cite{reisenegger2008a} and references therein), the considered
stage will be short. The secondary component does not have
sufficient time to undergo substantial evolution (the
characteristic time of the decay for the magnetar field lies in
the interval from several thousand years to several tens of
thousand years). Possible configurations can be identified in
which some feature of a magnetar, such as its strong magnetic
field, will be manifest in a critical way.

One example of a close binary system with a magnetar may be
provided by the central object in the supernova remnant RCW 103
\cite{pizzolato2008a}. This object displays variability with a
period of 6.7 h. One possible interpretation of the observations
is that the secondary is situated within the magnetosphere of the
magnetar \cite{pizzolato2008a,popov2006b}. In this case, 6.7 h is
the orbital period of the system. The binary is similar to polars,
in which the compact object is a white dwarf whose magnetic moment
ia approximately equal to that of the magnetar. In the
classification suggested by Shvartsman and Lipunov
\cite{lipunov1987a}, such systems are called {\it magnetors}. We
are planning to estimate the number of such objects in the Galaxy.

\section{Conclusion}

We have considered the hypothesis that GRBs and magnetars
originate during the evolution of massive stars in close binary
systems, which spins up the core of the pre-supernova. The
statistics for the expected rate of formation of magnetars are in
satisfactory agreement with observational estimates. However, this
scenario predicts a large fraction of binary magnetars, whereas
all known magnetars or magnetar candidates are single. This
problem may be solved by introducing an additional component of
the kick velocity acquired during the formation of the neutron
star, perpendicular to the orbital plane (i.e., along the
direction of the magnetic-dipole axis of the newborn compact
object), and requiring that the magnitude of the kick not be small
($\lesssim 400$ km/s). The presence of a moderately strong stellar
wind (evolutionary scenario C) also promotes a decrease in the
predicted binarity of potential magnetars; the requirement that
the kick velocity be high ismandatory, but it should be
preeminently directed perpendicular to the orbital plane of a
system.

\section*{Acknowledgments}

In our study, the population synthesis of close binaries was
carried out using the ``Scenario Machine'' code developed by V.M.
Lipunov, K.A. Postnov, and M.E. Prokhorov \cite{Lipunov1996} in
the Department of Relativistic Astrophysics of the Sternberg State
Astronomical Institute, Lomonosov State University, Moscow. This
work was supported by the Program of Support for Leading
Scientific Schools of the Russian Federation (grant no.
NSh-1685.2008.2), the Analytical Departmental Targetted Program
``The Development of the Science Potential of Higher Education''
(grant no. RNP 2.1.1.5940), the INTAS Foundation (grant no.
06-1000014-5706), and the Russian Foundation for Basic Research
(grant no. 07-02-00961). The authors thank Professor A.V. Tutukov
for discussions and useful remarks.

\end{document}